\documentclass[onecolumn,noshowpacs,preprintnumbers,amsmath,amssymb]{revtex4}
\usepackage{amsfonts}
\usepackage{amsmath}
\usepackage{ bbold }
\usepackage{graphicx}
\def \ee{\end{equation}}
\def \be{\begin{equation}}
\def \bea{\begin{eqnarray}}
\def \eea{\end{eqnarray}}
\def\lsim{\lower.5ex\hbox{$\; \buildrel < \over \sim \;$}}
\def\gsim{\lower.5ex\hbox{$\; \buildrel > \over \sim \;$}}

{\catcode`\|=\active\gdef\Braket#1{\left<\mathcode`\|"8000\let|\bravert {#1}\right>}}
\def\bravert{\egroup\,\vrule\,\bgroup}
\begin{document}

\title{A comment on the OPERA result and CPT}

\author{Benjamin Koch}
\affiliation{Pontificia Universidad Cat\'olica de Chile.\\ Av. Vicu\~na Mackenna
4860. Macul.\\ Santiago de Chile.}

\date{\today}
\email{bkoch@fis.puc.cl}

\begin{abstract}
We consider the possibility that the superluminal
neutrino propagation
reported by the OPERA collaboration 
originates from a violation of CPT.
On this basis we compare our actual knowledge 
concerning the CPT theorem to the nuclear 
reaction chain between CNGS and OPERA.
\end{abstract}

\maketitle
A recent experimental result from the OPERA collaboration
\cite{OPERA:2011zb} is that neutrinos 
of energies around $\langle E_\nu \rangle=17$~GeV 
seem to travel at velocities
of $v_\nu\approx 1.00002 c$. 
If this observation is confirmed it
 is surprising in various ways: 
 First,  it implies a violation of 
 conventional Lorentz variance at the order of 
$\epsilon_\Lambda=(v_\nu-c)/c=(2.48 \pm
0.58)\times 10^{-5}$.
Second, if the violation would originate
from quantum gravitational effects one would expect
it to be suppressed by the Planck mass $m_\nu/M_{Pl}\sim 10^{-28}$.
In this sense the reported effect is many orders
of magnitude larger than predicted by quantum
gravity of approaches.
Third, if one tries to accommodate this effect
in an extended theory of relativity
that by construction allows for superluminal
propagation, the observed velocity difference
appears highly fine tuned to a value much less than
one but different from zero. Thus, from this perspective
the value of $\epsilon_\Lambda$ is much smaller than expected.
Therefore, the value of $\epsilon_\Lambda$ is
unnatural for both, quantum gravity and extended Lorentz
approaches. 
Estimating the mass scale corresponding to this effect by
$\epsilon_\Lambda \sim m_\nu/M_X$ one finds that the scale $M_X$
should actually lie in the MeV range.
The violation of CP in this energy regime 
is well studied within the standard model~\cite{ManyCP}, therefore
it is natural to consider also the possibility of a violation
of CPT~\cite{CPT}. Since it has been shown that a violation of CPT implies a
violation of
Lorentz invariance~\cite{Greenberg:2002uu} this fact could
be relevant for an understanding of the origin and value of~$\epsilon_\Lambda$.

The idea of this comment is to study possible CPT \cite{CPT} violations  as
source of LI violations.
Please note that
a violation of CPT implies a violation of
Lorentz invariance~\cite{Greenberg:2002uu} while the inverse statement is not
necessarily true.
In order to obtain an estimate for the  value 
of CPT violation one has to consider the total nuclear reaction chain
between CNGS and OPERA.

This reaction chain $p \rightarrow (\mbox{mesons} +\dots) \rightarrow (\nu
+\dots)$
consists basically of three stages of propagation and two 
intermediate stages of conversion.
It is hard to imagine that the conversions (decays)
produce a Lorentz violating effect that results in an advance
of $\delta l =18$ meters. Also the velocity of the proton is known and measured
to high precision leaving only the stages of meson and neutrino
propagation as possible sources of Lorentz violation.
Estimating the amount of Lorentz violation due to
neutral Kaons 
\cite{Angelopoulos:2003hm,DiDomenico:2009zz} one gets
$\epsilon_\Lambda^{K_0} \approx Re(\delta^{K_0}_{CPT}) \quad
\mbox{with}
\quad 0.2\times 10^{-4}\le Re(\delta^{K_0}_{CPT}) \le 4.8\times
10^{-4}$, which would be in agreement with the
OPERA value. However, such an estimate is too naive. For example one knows
that the relevant mesonic multiplicity consists of 99 percent pions
and only about one percent neutral Kaons \cite{Bleicher:1999xi}.
Although it is sufficient to have only a fraction 
of the events with superluminal velocities \cite{Winter:2011zf},
it is clear that this fraction should exceed one percent.

There are also good experimental bounds on the 
amount CPT violation in pions \cite{PDG}.
However, the limit on CPT violation in pions 
measured at $\sim 0.3$~GeV \cite{Greenberg:1969da} is  $\Delta \tau_\pi 
= \epsilon_\pi \tau_\pi
\approx
5\cdot 10^{-4} \tau_\pi$.
By using the corresponding Lorentz $\gamma$ factor,
the corresponding limit on the advance within the lifetime of a
pion is
$\delta l \le \Delta \tau_\pi c \gamma \approx 0.5$~m,
which is a factor of thirtysix smaller, than the advance of $\delta l=18$~m
needed to explain the OPERA data.
Of course one has to notice that this comparison involves
different energy scales ($0.3$~GeV and $17$~GeV).
Considering a cinematic explanation of this discrepancy
already a linear energy dependence $\delta_{CPT}^\pi$
would be sufficient to get an agreement between
$\epsilon_\pi|_{CPT}$ and $\epsilon_\Lambda|_{OPERA}$.

The remaining possibility is the propagation of the neutrino.
It is known that the CPT violating terms in a
renormalizable Lorentz violating standard model
extension \cite{Kostelecky:1997mh,Diaz:2010ft}
can not generate superluminal neutrino velocities.
This implies that apparently the only realization of this idea
in the neutrino sector will imply non-renormalizable terms.
Unfortunately there is basically no model independent data
on CPT violation in the neutrino sector. Therefore,
the general comment has to end at this point.

As continuation a special toy model 
for CPT violation in the neutrino sector will be mentioned.
For other models and explanations the reader is referred to 
\cite{Alfaro:2004aa,Pas:2005rb,Dent:2007rk,Klinkhamer:2011mf,Wang:2011sz,
Wang:2011zk,Alfaro:2011sp,Giacosa:2011tp}.
A possible toy model for this idea can be constructed
in the spirit of \cite{Barenboim:2001ac} where a mass splitting 
between neutrino mass $m$ and anti-neutrino mass 
$\tilde m$ is the source of CPT violation.
The hamiltonian of a fermionic lagrangian is
\be
:H_0:=\int \frac{d^3p}{(2\pi)^3}\sum_s 
\left( 
({\bf p}^2+m^2)a_{{\bf p}}^{s\dagger}a_{{\bf p}}^s+
({\bf p}^2+\tilde m^2)b_{\bf p}^{s\dagger}b_{\bf p}^s
\right)
\ee
Since the CPT operator converts the operators $a_{\bf p}^s$ and $b_{\bf p}^s$
into each other, any difference between $m^2$ and $\tilde m^2$
immediately generates a violation of CPT.
This idea has already shown its utility in the context
of explaining neutrino oscillations and the LSND results
\cite{Barenboim:2004wu,Mavromatos:2011ur}.
As a slight modification one can consider that from
a certain energy scale $M_X$ on a new interaction
with width $\Gamma$ introduces
a momentum dependent splitting of the type
\be\label{split}
\tilde m^2- m^2=2 \epsilon \cdot M_X^2 \left({\mbox{Erf}}(({\bf
p}-M_X)/\Gamma)-1\right) \quad.
\ee
Due to the error function the splitting
is just generated above the momentum scale $M_X$,
while the CPT is conserved below $M_X$.

\begin{figure}[*h!t!p!]
\includegraphics[width=0.5\textwidth]{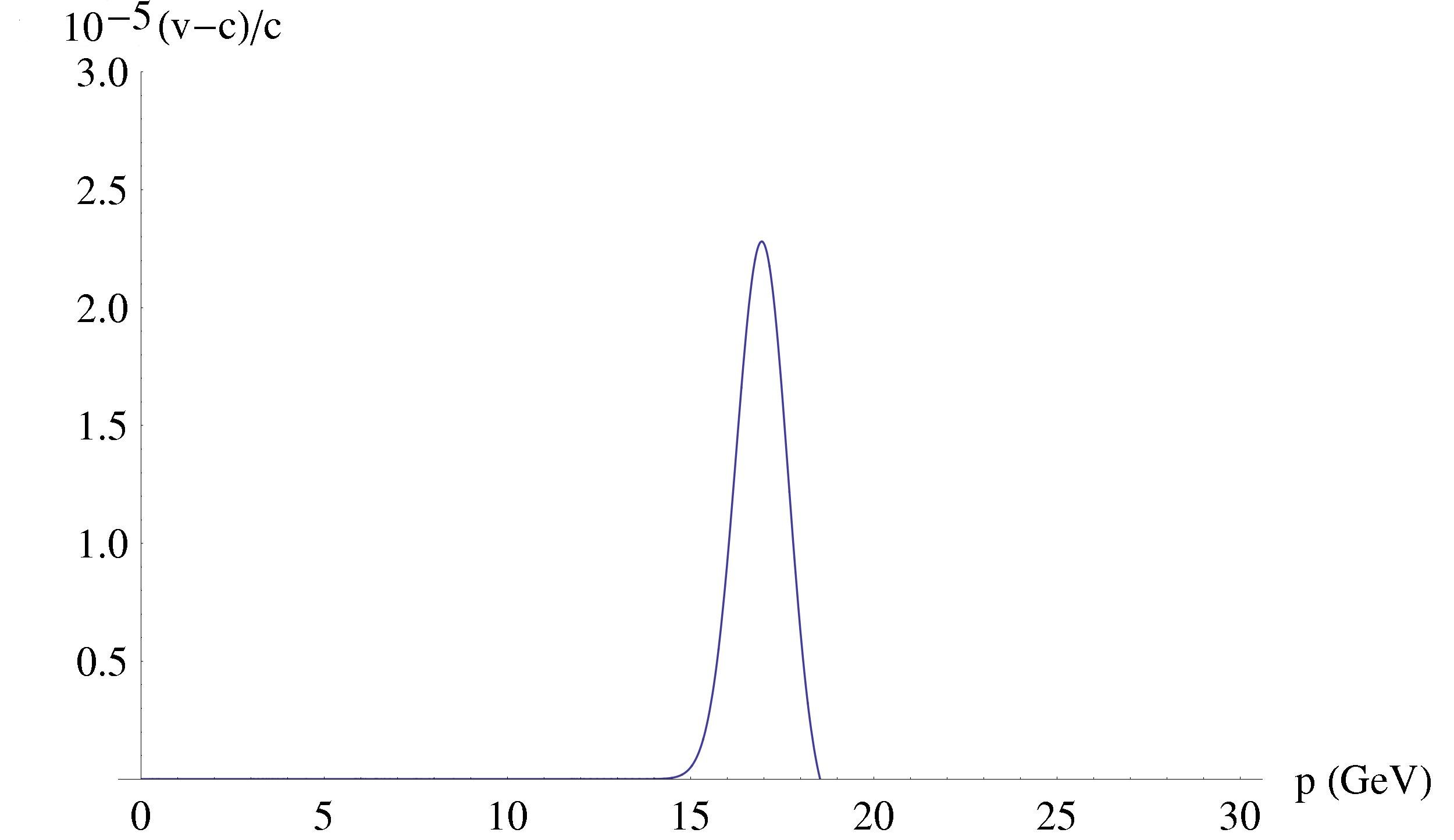}
\caption{Example for (\ref{deltav}) with $m\ll M_X=17$ GeV, $\Gamma=1$ GeV,
and $\epsilon=2.5\cdot 10^{-5}$}
\label{secint}
\end{figure}

The violation parameter $\epsilon$ has mass dimension
zero and it is assumed that
$\epsilon \ll 1$.
The dispersion relation of the neutrino would then be
\be
E^2={\bf p}^2+\epsilon \left({\mbox{Erf}}(({\bf
p}-M_X)/\Gamma)-1\right)+m^2\quad.
\ee
Thus, the group velocity for neutrinos with mass $m$ is
\be\label{deltav}
\frac{v}{c}=\partial_p E 
=1+\epsilon \frac{M_X^2}{\sqrt{\pi}\Gamma  |{\bf p}|} \exp \left(-(|{\bf
p}|-M_X)^2/\Gamma^2\right)
+{\mathcal O}
\left(\epsilon \frac{m^2}{{\bf p}^2},\epsilon^2,\frac{m^4}{{\bf p}^4}\right)
\ee
where an expansion for large momentum and small CPT violating
parameter was used.
One observes that around $M_X$, the enhancement is $\Delta v/c\sim \epsilon
p/\Gamma$, whereas for larger values of p it is $\sim \epsilon$.
Thus, the relation (\ref{deltav}) has the effect
that around the scale $M_X$ a large enhance
of group velocity occurs, whereas before and after,
this effect is zero or negligible  as it is shown in
figure (\ref{secint}).
Such a cinematic behavior could
explain the absence of significant superluminal velocities 
of neutrinos from the supernova 1987a.
This energy dependence would also be essentially
different from the power law deviations and respective
bounds studied in \cite{Giudice:2011mm,Alexandre:2011bu}.
As explained in \cite{Murayama:2000hm,Barenboim:2004wu}
CPT violating mass differences could also
have effects on special types of neutrino oscillation
experiments \cite{Aguilar:2001ty}.
Thus, any model like (\ref{split}) should
be confronted with all available neutrino data
\cite{Fargion:2011hd,Bi:2011nd,Dass:2011yj},
which goes beyond the scope of this comment.

Although the relation (\ref{split}) seems to work potentially
well, the point of this paper is not to put forward
a specific model. The idea is to emphasize
the possible connection between a tiny CPT violation $\epsilon$
and the small deviation from $c$ reported by OPERA.
It is found that for a constant
CPT violation in pions $\epsilon_\pi$ is not sufficient
to be the source of the observed superluminal propagation.
This restriction, does however not apply if one considers
a $\epsilon_\pi$ that increases linearly with the energy.
Additionally, also
the propagation of the neutrino has the potential
of describing such a sizable observational effect.
Model independent tests of CPT violation in neutrinos
are however missing.
For both, the neutrino and the pion sector one would have to invoke some kind of
energy dependent CPT violation in order to
evade contradiction between different experiments. 

Many thanks to M.A. Diaz, J.S. Diaz, 
and J. Gamboa for valuable hints. The work of B. K. was supported by CONICYT
project PBCTNRO PSD-73. 


\end{document}